# Internet of Things: Infrastructure, Architecture, Security and Privacy

Zainab Alansari
*Department of Computer Studies*
*AMA International University*
*BAHRAIN, Bahrain*
*University of Malaya, Malaysia*
zeinab@amaiu.edu.bh

Nor Badrul Anuar
*Department of Computer System & Technology*
*University of Malaya*
Kuala Lumpur, Malaysia
badrul@um.edu.my

Amirrudin Kamsin
*Department of Computer System & Technology*
*University of Malaya*
Kuala Lumpur, Malaysia
amir@um.edu.my

Mohammad Riyaz Belgaum
*Department of Computer Studies*
*AMA International University*
Salmabad, Bahrain
bmdriyaz@amaiu.edu.bh

Jawdat Alshaer
*Department of Computer Information System*
*Al-balqa Applied University*
Salt, Jordan
jawdat_alshaer@bau.edu.jo

Safeeullah Soomro
*Department of Computer Studies*
*AMA International University,*
Salmabad, Bahrain
s.soomro@amaiu.edu.bh

Mahdi H. Miraz
*Centre for Financial Regulation and Economic Development (CFRED)*
*The Chineese University of Hong Kong*
Sha Tin, Hong Kong
m.miraz@cuhk.edu.hk

*Abstract*—**Internet of Things (IoT) is one of the emerging technologies of this century and its various aspects, such as the Infrastructure, Security, Architecture and Privacy, play an important role in shaping the future of the digitalised world. Internet of Things devices are connected through sensors which have significant impacts on the data and its security. In this research, we used IoT five layered architecture of the Internet of Things to address the security and private issues of IoT enabled services and applications. Furthermore, a detailed survey on Internet of Things infrastructure, architecture, security, and privacy of the heterogeneous objects were presented. The paper identifies the major challenge in the field of IoT; one of them is to secure the data while accessing the objects through sensing machines. This research advocates the importance of securing the IoT ecosystem at each layer resulting in an enhanced overall security of the connected devices as well as the data generated. Thus, this paper put forwards a security model to be utilised by the researchers, manufacturers and developers of IoT devices, applications and services.**

## I. Introduction

The term "Internet of Things" was first used by Kevin Ashton in 1999 [1], and was first introduced to the (Massachusetts Institute of Technology) [2] when the world was described in a vision where all our personal and working devices including inanimate objects, have not only digital identity but potential processing ability allowing central computer system to organize and manage them.

The concept of the Internet of Things was introduced by a researcher in the RFID Development Committee in 2000 [3]. He claimed the possibility of discovering information about a tagged object by searching in a specific Internet address in the contents of a database [4]. Since then, the term Internet of Things has been tied to everyday detected, located, addressed and controlled objects via the Internet, either through Radio Frequency Identification (RFID) [5], wireless network [6], Wide Area Network (WAN) or other tools) [7]. Objects not only include electrical appliances, but also other non-electrical ones such as food, clothing, materials, parts and so on [1].

Many definitions of the Internet of Things by various research associations have been expressed based on their attitude towards the strengths of the idea [8]. The reason for this ambiguity is the name of the concept "Internet of Things" [9] which is composed of two words; the first word emphasizes the networking view of this concept, while the second word emphasizes the move to the general objects that are in a common package. Whether to consider objects that are Internet-oriented, or entity-oriented is left on the type of the use and the environment where it is used [10]. The Internet of Things will lead to changes in stakeholders, business contracts, research and standards [11]. In fact, the Internet of Things means a global network of related objects, each of which has its own address, that is in accordance with standardized contracts. Unique addressing, displaying and storing information exchanged in objects is a challenging topic that incorporates the view of the "Internet of Things [12]."

### A. *Internet of Things Object-Oriented Perspectives*

The use of radio frequencies is the first and easiest way to realize this view. The Auto-ID Association (one of the global pioneers in developing the Internet of Things) considers the term Internet of Things to be a vast network of signs of radio frequency and sensor technology [13]. The goal of establishing these types of definitions was to achieve a common architecture on the Internet of Things based on the unique electronic object's code. The electronic code of an object was actually designed to support the expansion use of radio frequency signals in modern business networks and to create the required standards for these networks. These standards are designed to improve the visibility of objects such as tracking, providing state awareness and determining the location of objects [14]. Radio frequency signals are still highly utilised in the Internet because of its low cost and strong support as a leading technology in this view [15]. However, technologies such as Near-field communication (NFC) [16] and Wireless Sensor Networks (WSN) and wireless devices which are compatible with radio frequency signals are also applicable. Considerable amount of research is being conducted in developing proper framework for wireless sensor networks called Wireless Sensor Sensing. Many of the research focusing on Internet of Things identified that the concept of the Internet of objects places more emphasis on entities, these objects playing the leading role [17]. In fact, this definition come up with the theory of real-world intelligent objects implementation [18]. The above components should have the ability to communicate, to display autonomous behaviour, in addition to being commonly used with wireless sensors.

### B. *Internet of Things Internet-Oriented Perspectives*

In 2008, the 25 largest companies in the world agreed on a protocol called "IP for Smart Objects (IPSO)" network communication [3]. Based on the protocol, large volumes of devices on which embedded systems are running can communicate with each other. The protocol guarantees the use of IP for the development of the "Internet of Things." To reach the IPSO, the IEEE community has included the IEEE 802.15.4 standard in the IP architecture.

### C. *Internet of Things Semantic-Oriented Perspective*

The number of components or devices that forms the Internet will be increased in the future, rising the issues like how to display, store, communicate, search and organize information, resulting in many new challenges. In this perspective, semantic technologies can play a key role. In fact, these are suitable solutions for describing the entities, processing events generated on the Internet of Things, semantic environments, and architectures [7]. These perspectives are adapted to Internet requirement of objects, as well as to the scalable storage and communication structures [13]. Fig. 1. displays the three perspectives to the Internet of Things paradigm.

Over the past decade the advancements in the field of Internet of Things has been accelerating with a greater speed to provide a comfortable and luxurious style in the lives of the users in the society. Its use in different areas is taking the users to a different era where the users too did not expect [19]. Internets of Things are the objects which are uniquely identified and are connected in an internet like structure. Capturing Internet of Things is nothing but having a full control on the data coming from different devices which are already connected and more devices to get connected as part of its elasticity property [6].

In this paper, we review the Internet of Things concepts and definitions and define the three perspectives of Internet of Things paradigm. The structure of the rest of the paper is as follows: section two gives a broad explanation of Internet of Things infrastructures while section three discusses the Internet of Things communications. Furthermore, section four illustrates the Internet of Things Architectures and finally in section five we present a detailed comparison as well as analysis of Internet of Things Security and privacy.

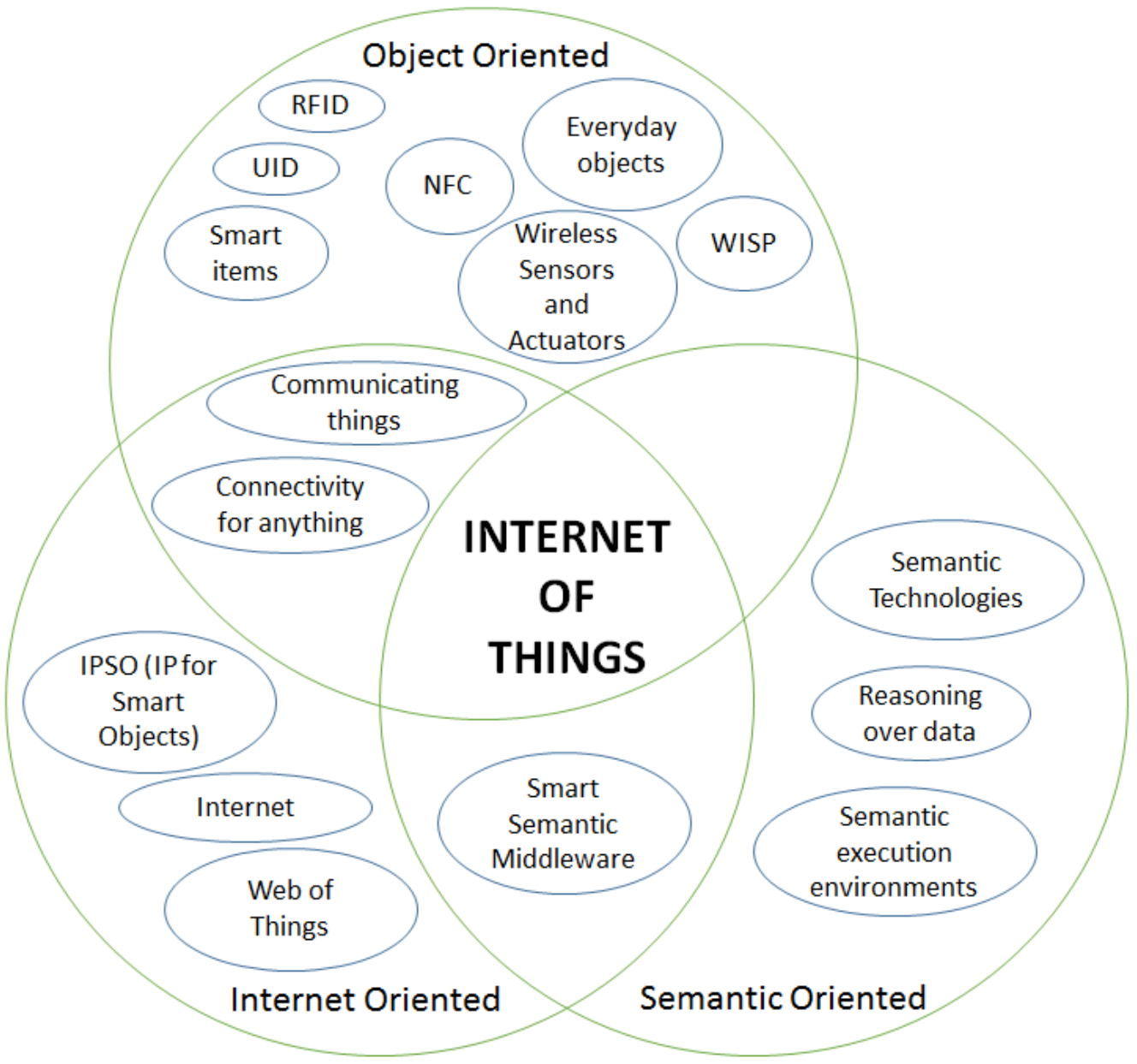

Fig. 1. The three perspectives to the Internet of Things paradigm.

## II. Internet of Things Infrastructures

Scalability, commitment, extendibility, and interoperability between heterogeneous objects and their environment are the key requirements for Internet of Things [5]. In order to ensure open source space for developers and solution providers, the Internet of Things requires an open source architecture to maximize interoperability between heterogeneous systems and distributed resources, including providers and consumers of information and services (including humans, software, smart objects [20], or other devices).

At the heart of the Internet of Things, millions of devices send their data to centralized systems [21]: data that usually does not have any similarity amongst themselves. In other words, while some Internet of Things devices may collect data related to humidity and temperature, others may collect data related to the people's location or their daily activities. Eventually the data is sent for analysis to cloud servers [5] or other Internet of Things devices [22]. Therefore, the existence of a reliable and high-speed connection plays a key role in communicating between Internet of Things devices [16]. Wi-Fi, Bluetooth, RFID, NFC, Zigbee, etc [9] are a few options that are available for Internet of Things devices. Each of these options depends on a variety of factors, such as sensor density, speed, environmental factors (type of environmental equipment: concrete, wood, metal) and the extent to which Internet of Things devices are used [23].

Conceptually, the Internet of Things is based on three critical principles that are related to objects intelligence: 1) the automatic identification of smart objects [20], 2) the ability of intelligent objects to communicate with their surroundings [24], and 3) the ability of interactions between objects themselves.

Objects applying these principals in the main network send information to the users (companies, end-users, service organizations, government, etc) [25]. In empowering such an approach [16], the main challenge is the development of web platforms, web-based software, service-oriented architectures and intelligent technologies. In this intelligent technology, objects play a key role, and combining them with a variety of technologies, such as RFID, sensors and microchips, requires specific rules and standards.

The main focus is on the Internet of Things and information, rather than point-to-point communications [26]. This fact can be adapted to the principles and architecture of the content-oriented networks [4]. While the Internet of Things perspective represents significant advances in some areas of the IT field, its realization requires a gradual process that begins with existing technologies and applications. In particular, the Internet of Things starts with identifying technologies, such as RFID (widely used in many cases), which have been long used in the past, and simultaneously, on the development path, of the capabilities of the wireless sensor network technology (as a tool for data collection and service-oriented architecture (as a software for developing Web-based services).

The Internet of Things is not a unique technology. It is a set of empowering and ideological technologies. One aspect of empowerment is to shrink electronic components and reduce their prices. Wireless transmission technologies provide easy communication for the objects. On the other hand, the advancements in the domain of sensor technology produce smaller and cheaper sensors that consume less energy and provide positioning (GPS) technology and information about the absolute or relative position of the components. All of the above points define the potential for the expansion of the Internet of Things [6]. When extending the Internet of Things; the most important factor is the implementation of an integrated standard for the different infrastructures of Internet of Things. Moreover, how to use, the typical data formats is followed by a series of rules to create interoperability between different objects. Logically, the network should be organized in such a way that the IP packets automatically find their routes [27].

## III. Internet of Things Communications

Generally, there are three different communication modes for the objects: 1- Object-to-human 2- Object-to-object. 3. Object-to-Human connection.

Object-to-human communications allow the remote control on objects by the human. Reports generated by such communication paradigm typically include: location, status, and sensor information [28]. Object-to-object communication involves a number of technologies and software, in which everyday objects communicate with other objects without human interference [29]. These objects can monitor other objects and perform actions to correct the situation or send messages to the human if deemed to be necessary [30].

Nowadays, widely used objects contain microcontrollers, where wireless interconnections are their cores. These microcontrollers have been integrated to the memory, software, drivers and interfaces of the sensors by adding a network interface, individuals and objects can monitor and control other objects through the Internet. Software that is deployed on servers or objects can perform their functionality with or without human intervention. The combination of microcontrollers, memory, software, interface drivers, as well as sensors, intrinsically enables the network connectivity [31]. The Internet of Things consists of everyday objects that have at least one electronic identifier. It's estimated that on average each person is surrounded by a huge number of objects [32]. If all objects are identifiable, the Internet of Things may have 50,000 to 100,000 billion members [13]. There are various methods to deal with this overload problem. The old and most widely used Internet Protocol Version 4 (ipv4) with 32 bits only has the ability to create 4 billion addresses. As a result, the use of IPv4 cannot be continued for long. The new Internet Protocol version 6 (IPv6) provides 128-bit which is capable of addressing the increasing IDs [3].

## IV. Internet of Things Architecture

Internet of Things, as a hot topic in the field of computer networks and wireless sensor networks, requires a standard architecture to provide a competitive environment for business organizations competitors in order to enhance their quality. In addition, detailed evaluation of the traditional Internet architectures needs to be performed to measure its capability to meet the challenges of the Internet of Things. The Internet of Things connects a large number of heterogeneous objects through the Internet. Hence, there should be built in flexible layered architecture [11]. Different architectures can be deployed for the Internet of Things, such as three-layer architecture, middleware architecture, service-based architecture and five-layer architecture. Hence The common architectures of the Internet of Things, which are Perception Layer, Network Layer, Middleware Layer, Application Layer and Business Layer, are presented in Fig. 2.

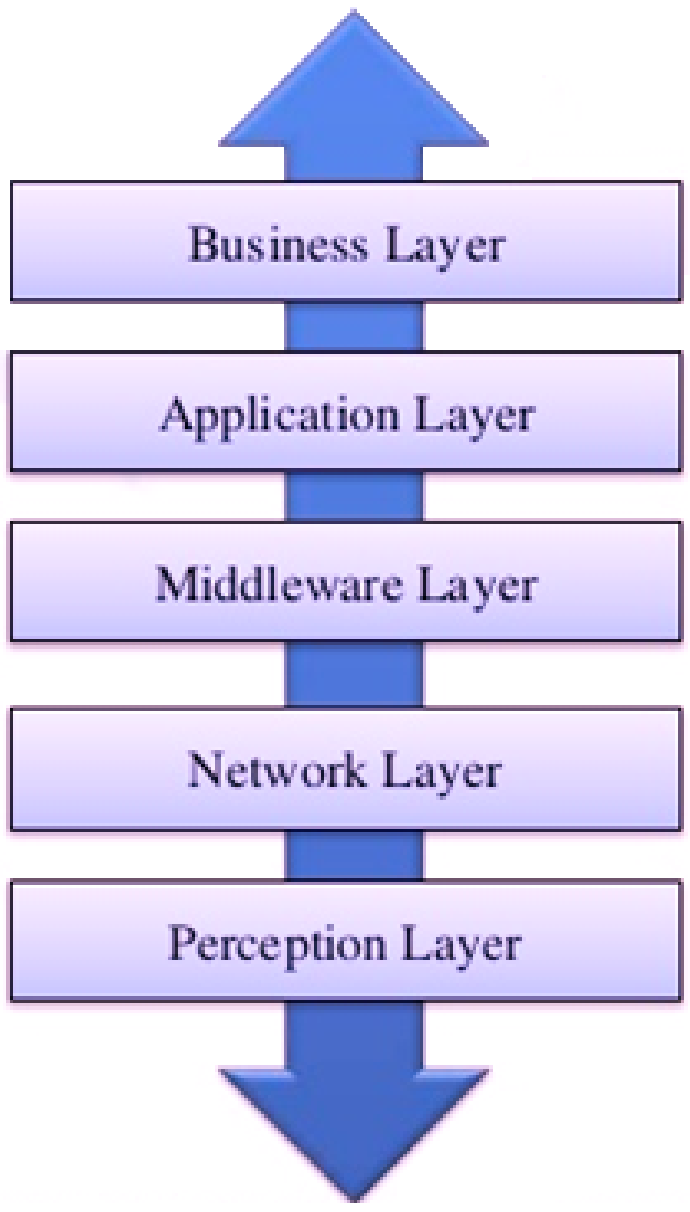

Fig. 2. The five-layer Architecture of Internet of Things.

### A. *Perception Layer*

The first layer is the perception or objects layer. In this layer, the physical sensors collect and process the information, digitizes the collected data and them through the secure channel to the Network layer. Perception layer includes sensors and stimuli for various tasks of detecting, measuring and reading such as location, temperature, weight, movement, vibration and humidity. In short, the journey of the massive data generated from different entities and loaded in the Internet begins with the perception layer [33].

### B. *Network Layer*

On this layer, the produced data from the perception layer are sent through the secure channel to the Middleware Layer [34]. Data can be transmitted using technologies such as RFID, Wireless communication systems, Global System for Mobile communications (GSM), Wi-Fi, Bluetooth, etc [7]. The data collected from the objects can be stored and processed using cloud services and data management processes are controlled in this layer [35].

### C. *Middleware Layer*

The Middleware Layer from the layered architecture of Internet of Things divides a service by name and address with the information received from the requester. This layer allows developers to work with heterogeneous objects regardless of the hardware platform [36].

This layer processes the received data, defines and delivers the required service. The middleware is fundamentally a software layer between the hardware components and applications in the Internet of Things [1]. In other words, a set of sub-layers are used between the level of application software and technologies. In recent years, the middleware played a major and important role in developing new services and combining the benefits of past technologies to achieve the new ones [14]. There are several reasons for the need of having an interface in Internet of Things. The first reason is that the description and implementation of common standards are very difficult considering the fact that there are various technologies on the Internet of Things. Hence, the Internet of things must be connected to heterogeneous computers, the middleware acts as a link between different technologies. The second reason is that there is a demand for separating the application layers in different domains. To mitigate this problem, the middleware provides applications for graphical interfaces for physical layer communication services. The middleware also hides the unnecessary details and variety of physical layer technologies to simplify the overall process, especially to the users. In other words, information gathered by intelligent devices are being sent to a comprehensive data collection system through various networks that are nowadays the mainstream of the IT industry [13]. The user's information is provided without any complexity through middleware, which is the task of data synchronization and resource combinations.

### D. *Application Layer*

This layer is deployed upon client request, fully dependent on the functionality values of entities, examples include: temperature and humidity measurements. The importance of this layer is due to the fact that it can provide high-quality intelligent services to meet various customer requirements [37].

### E. *Business Layer*

This layer manages all the system activities and the services. The responsibility of this layer is to build a business model, graphs, flowcharts etc. based on the data received from the application layer. This layer also designs, analyses, implements, evaluates, monitors and develops elements of the Internet of Things. In addition, managing and monitoring the underlying layers is also possible from this layer [38]. Moreover, it compares the output of each layer with the desired output to improve services and maintain the privacy of its users.

In general, in the architecture of the five layers of the Internet of Things, objects are in the first layer with the help of a set of sensors and network facilities and data are collected and aggregated to continue with the help of the second layer. Collected data are available to the third layer for management and processing. The result of the services provided by the three layers are software in the fourth layer, which provides the necessary benefits of the Internet of Things in realizing the needs of a particular industry or a set of different companies. The companies that are going to enter the market are considering only the familiarity and existing capacities of Internet of Things, ignoring the opportunities and threats.

## V. Internet of Things Security and Privacy

One of the major challenges that must be addressed in order to import the Internet of Things into the real world is security. Internet of Things is supposed to deal with a population of billions of objects that interact with each other and with other things, such as humans or virtual entities. All of these interactions need to be protected in some way, including protecting information and providing services to all relevant actors, as well as limiting the number of incidents affecting the entire Internet of Things. However, Internet of Things protection is a complicated and difficult task. The number of attacks available by malicious attackers may conflict with the main trends in the Internet of Things [39] due to universal connectivity (access for anyone) and accessibility (access to any location at any time). Threats that can affect Internet of Things are numerous, such as attacks on various communication channels, physical threats, denial of service and identity theft. Moreover, multiple heterogeneous entities can exchange information in different fields, securing of which requires more complex design and implementation of efficient, consistent, and scalable security mechanisms [34].

The main communication channel of the Internet of Things is the Internet. Therefore, Internet applications must be built to protect against both active and passive attackers. In addition to the Internet of Things security, the Internet of Things infrastructure must provide intranet security, data security, software security, hardware security and physical security. The popularity of Internet of Things will attract cybercriminals to attack the data points, transmitting commands, network points and entries in the first stage. So specific protection should be provided for these points. Considering the influence of Internet of Things in all aspects of life and the threat of individuals by malware, the use of secure architecture to deal with these threats is highly required [28]. One of the mechanisms for creating security for the Internet of Things is to build right secured Internet of Things architecture on all the layers.

Security and privacy are the two important and complex challenges of the Internet of Thing. Breaching of security includes unauthorized access to information and attacks that cause a definite physical disruption to the service's accessibility [39]. While digital citizens are becoming increasingly instrumental providing with the data related to their locations and activities, privacy seems to disappear. Privacy protection systems collect data and send emergency responses when technological challenges come together with ongoing security challenges. This is necessary for a smart city that we want to live in. Security and privacy are widely known issues of the Internet of Things [9]. On the other hand, the confidentiality and integrity of the data transmitted and stored must be guaranteed, and authentication and licensing mechanisms should be provided to prevent inappropriate access of users or unauthorized devices. Furthermore, the privacy of users, such as the ability to support data protection and the anonymity of users, should be considered as a fundamental aspect, especially in providing sensitive or personal information. Moreover, data quality is also a necessary requirement for the implementation of objects in the scale of Internet of Things services [12]. Since, in some cases, wrong or missing values may have a significant effect on the actions or decisions of the Internet of Things, the information provided must be accurate, timely and complete [40]. In fact, as services and applications on the Internet of Things use different data sources, the user (or software) must be aware of the level of security and quality of available data in order to make decisions about his or her use.

Internet of Things is becoming a key element of the future Internet and a vital national and international infrastructure [14]. Under these circumstances, providing sufficient security for the Internet of Things is becoming increasingly important. Large-scale applications and Internet-based services are increasingly vulnerable to any disturbance, attack, or theft of information [27].

On the other hand, considering privacy rights is a prerequisite for building trust on the Internet. The Internet of Things is often referred to as a large network of sensor-powered devices designed to collect data from their surroundings (including data related to individuals). The collection and use of Internet of Things data requires confidentiality. It should be noted that combining Internet of Things data can greatly affect confidentiality [11]. For example, a toothbrush connected to the Internet may have an application that collects and sends information (harmless) related to its behaviour [7]. But when these data are combined with the data reported by the user's refrigerator in relation to what the person is eating and the data reported by his health monitoring device, a much more intuitive picture and a more confidential description of the general health will be released.

Today, our daily activities depend on equipment or systems that have Internet connectivity. Many manufacturers also offer their products with the ability to connect to the Internet [40]. Therefore, Internet of Things equipment needs to be updated day by day, and thus the security of these equipment becomes more intense. This will increase the level of dependence on Internet of Things equipment and Internet services providing penetration routes and access to equipment to the hackers [2]. Perhaps, we can easily turn off an internet-connected TV and thus turn off a cybercriminal environment, but it's not easy to incapacitate an intelligent power meter, traffic control system, or heart rate regulator installed on patient's body. Thus, the security of Internet of Things equipment and its services are considered to be the core issues in this area. In fact, there are various factors that influence the assessment of security risks and how to deal with them[18]. These factors include: accurate identification of current and future security risks, estimating financial and non-financial costs when these risks are implemented and estimating the cost of reducing security risks. Fig.3. Illustrates such multifaceted security requirements in the domain of the Internet of Things.

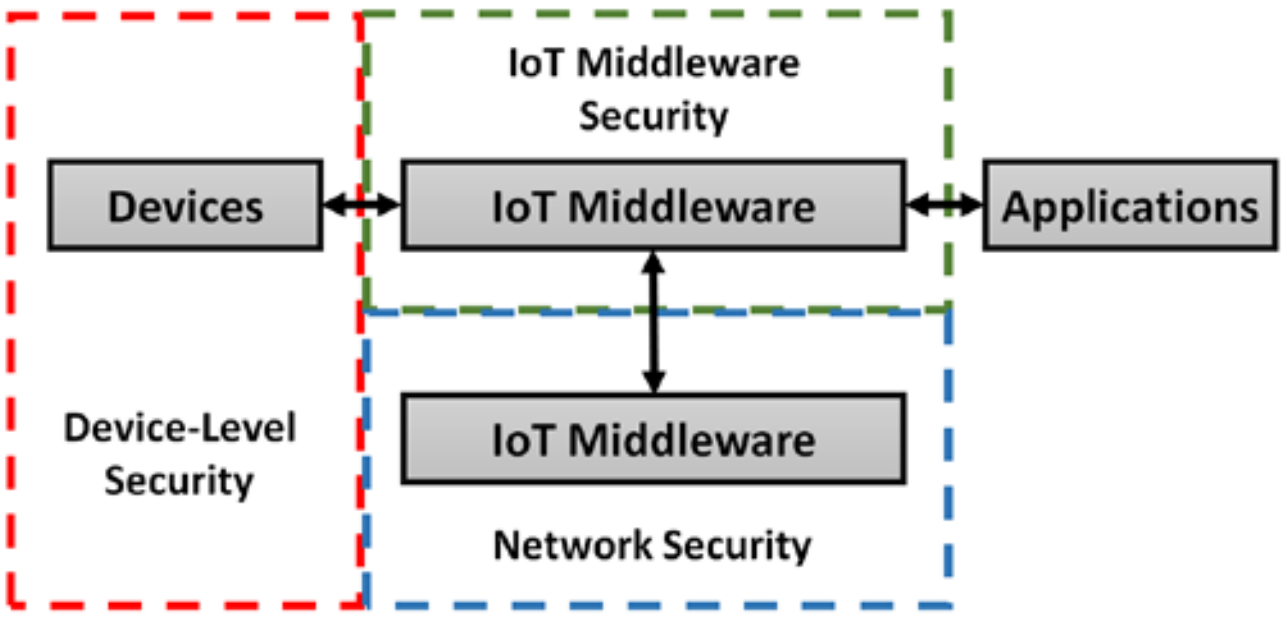

Fig. 3. Internet of Things Security.

## VI. CONCLUSION

In this paper, the layered architecture of Internet of Things was presented for the purpose of simplifying the theory and the practice concepts not only for Internet of Things as coming revolution but also as open security challenge. In this regard, five-layer architecture for the Internet of Things was introduced. All organizations that intend to use the potential of the Internet of Things, in the first stage, should develop a codified strategy in this area and should have a framework to handle the vulnerabilities arises. Drawing up a strategy focusing on the architecture mentioned can greatly increase the success of the implementation. In turn, it minimizes the security holes which will lead to secure realization of the Internet of Things applications. The major concern i.e. security is dominating over convenience in use of the Internet of Things devices or the network connecting these devices. Future research plans include the use of statistical fingerprinting [41] and Blockchain [42] technologies to enhance IoT ecosystem.

## ACKNOWLEDGMENT

This work is supported by CTRG Research Group of the College of Computer Studies, AMA International University BAHRAIN, Kingdom of Bahrain.